\documentclass[letterpaper,11pt]{article}
\pdfoutput=1 
\usepackage{jheppub} 
\usepackage[perpage]{footmisc}
\usepackage{amsmath,amsfonts,amssymb}
\usepackage{psfrag}
\usepackage{enumerate}
\usepackage{mathrsfs}
\usepackage{graphicx}
\usepackage{wrapfig}
\usepackage{xcolor}
\usepackage{caption}

\newcommand{\N}{{}^{i}\hspace{-1.1mm}N}
\newcommand{\Nj}{{}^{j}\hspace{-1.1mm}N}
\newcommand{\Nn}{{}^{0}\hspace{-1.1mm}N}
\newcommand{\Ne}{{}^{1}\hspace{-1.1mm}N}
\newcommand{\K}{{}^{i}\hspace{-1.1mm}K}
\newcommand{\Kj}{{}^{j}\hspace{-1.1mm}K}
\newcommand{\Kn}{{}^{0}\hspace{-1.1mm}K}
\newcommand{\Ke}{{}^{1}\hspace{-1.1mm}K}
 
\makeatletter
\def\@fpheader{\relax}
\makeatother

\begin{document}

\title{\centering Metric Perturbations of Extremal Surfaces}

\author{Benjamin Mosk}%\email{bmosk1@stanford.edu}
\affiliation{Stanford Institute for Theoretical Physics, Department of Physics, Stanford University\\
	Stanford, CA 94305, USA}
\emailAdd{bmosk1@stanford.edu, benjaminmosk@gmail.com}
\vspace{10pt}
%\begin{indented}
%	\item[]October 2017
%\end{indented}

\abstract{
Motivated by the HRRT-formula for holographic entanglement entropy, we consider the following question: what are the position and the surface area of extremal surfaces in a perturbed geometry, given their anchor on the asymptotic boundary? We derive explicit expressions for the change in position and surface area, thereby providing a closed form expression for the canonical energy. We find that a perturbation governed by some small parameter $\lambda$ yields an expansion of the surface area in terms of a highly non-local expression involving multiple integrals of geometric quantities over the original extremal surface.
}

\maketitle
%\tableofcontents
%\tableofcontents
\newpage

\section{Introduction}
The AdS/CFT-correspondence is the conjecture that under certain conditions, a conformal field theory (CFT) on a $d$-dimensional spacetime (``the boundary'') describes a theory of quantum gravity on a $(d+1)$-dimensional asymptotically Anti de Sitter (AdS) spacetime (``the bulk'') \cite{Maldacena:1997re}. 

The ``holographic dictionary'', the map between quantities in these dual theories, is a subject of continued study. An important identification was given by the HRRT-formula \cite{Ryu:2006bv,Hubeny:2007xt}, which states that the entanglement entropy of a sub-region of the CFT can be associated with the area of an extremal surface in the bulk.

In the AdS/CFT-correspondence, perturbations of the quantum state of the CFT can be associated with perturbations of the bulk geometry and bulk fields. The CFT-state is thought to describe a well defined classical bulk geometry only under certain conditions \cite{Botta-Cantcheff:2015sav,Christodoulou:2016nej}. The HRRT-formula was previously used to derive such a condition on the CFT \cite{Faulkner:2017tkh}. A deeper understanding of the relation between entropy in the CFT and the area operator in the bulk is expected to reveal more information about the following questions:
\begin{itemize}
	\renewcommand{\labelitemi}{\scriptsize$\blacksquare$}
	\item which CFTs are ``holographic''?
	\item which CFT-states are dual to well defined classical bulk geometries?
\end{itemize}

In this article, we obtain explicit expressions for the change of the position and surface area of extremal surfaces, due to perturbations of the bulk metric. 

A related geometric problem was previously discussed in the mathematics literature: what is the change of the area, when a surface is arbitrarily shifted away from extremality? At leading order in the ``shift'', the change of the area is given by an integral involving the Jacobi or stability operator \cite{colding1999minimal,calegari2007foliations}. Here, we consider a more complicated problem: first we perturb the metric, and subsequently we shift the surface such that it is extremal in the new, perturbed geometry. 

Previous work in this direction includes \cite{2008arXiv0802.2250A}, in which the metric perturbation of renormalized areas of extremal surfaces in $\mathbb{H}^3$ was studied (to second order). For certain black hole geometries, the corrections to holographic entanglement entropy were discussed in \cite{He:2014lfa} and \cite{Ghosh:2016fop} (also up to second order). Shortly after this article appeared, \cite{Ghosh:2017ygi} came out, of which the results partially overlap with this study. A different type of perturbations, those of the shape of the ``anchoring surface'' at the asymptotic boundary, were discussed in \cite{Faulkner:2015csl,Carmi:2017ezk}. In \cite{Bueno:2016gnv}, it was discussed how a number of problems can be addressed with the explicit perturbative expression for holographic entanglement entropy.

Our procedure allows for an explicit expansion of the area operator in terms of the metric perturbation, in a general gauge, and to arbitrary order. 

We also provide an iterative procedure to construct a diffeomorphism that brings the metric in the Hollands-Wald (HW) gauge \cite{Hollands:2012sf}, a gauge that was found to be useful in the context of the AdS/CFT-correspondence \cite{LashkariVanRaamsdonk2016,Beach:2016ocq,Faulkner:2017tkh}.

\section{Summary}
\subsection{Method}
The HRRT-formula \cite{Ryu:2006bv,Hubeny:2007xt} states that the entanglement entropy $S(B)$ of a (spatial) sub-region $B$ of the CFT, can be associated to the surface area of the bulk surface $\tilde{B}$, where $\tilde{B}$ has extremal area, and where $\tilde{B}$ is homological to the sub-region $B$ at the asymptotic boundary, by\footnote{At leading order in $G_N$, and in Planck units ($\hbar = c =1$).}
\begin{align}\label{eq:hrrt}
S(B) = \frac{A(\tilde{B})}{4G_N}.
\end{align}
There are corrections to this formula at sub-leading orders of $G_N$, that depend on other bulk fields, in addition to the bulk metric $g$ \cite{Faulkner:2013ana}. The HRRT-formula has been proved under mild assumptions \cite{Lewkowycz:2013nqa,Dong2016,Dong:2017xht}.

The quantum state of the CFT can be described by a density matrix $\rho$. We consider perturbations of a reference state $\rho_0$, typically the vacuum, which are governed by a small parameter $\lambda$:\footnote{Where the $\rho_i$ must be traceless and hermitian, for $i>0$.}
\begin{align}\label{eq:denspert}
\rho(\lambda) &= \rho_0 + \Delta \rho = \rho_0 + \lambda \rho_1 + \lambda^2\rho_2+\dots,
\end{align}
where $\rho_0$ is the density matrix associated with the background bulk geometry $g_0$.
When the perturbation of the state (\ref{eq:denspert}) corresponds to a classical bulk geometry, the metric can be expanded as 
\begin{align}\label{eq:metricpert}
g_{ab}(\lambda) &= g^0_{ab} + \Delta g_{ab} 
&= g^0_{ab}+ \lambda  g^1_{ab} + \lambda^2  g^2_{ab}+\dots 
\end{align}
where $\Delta g$ vanishes near the asymptotic boundary.
In this article we consider the extremal surface that ends on an arbitrary sub-region of the CFT on the asymptotic boundary and 
\begin{itemize}
	\renewcommand{\labelitemi}{\scriptsize$\blacksquare$}
	\item determine the location of the extremal surface in the perturbed geometry (\ref{eq:metricpert}) 
	\item derive a perturbative expansion for the surface area of the extremal surface in the perturbed geometry (\ref{eq:metricpert}).
\end{itemize}
In section (\ref{sec:var}) we use a variational method, where we 
\begin{enumerate}
	\item expand the embedding function $x^a(\alpha)$ as:
	\begin{align}\label{eq:positionpert}
	x^a(\alpha) &= x^a_0(\alpha) + \Delta x^a(\alpha) 
	= x^a_0(\alpha)+ \lambda x^a_1(\alpha) + \lambda^2x^a_2(\alpha)\dots
	\end{align}
	where $\{\alpha\}$ is a set of $(d-1)$ parameters for the extremal surface, and $x_0^a(\alpha)$ is the embedding function of the original surface $\tilde{B}$, which is extremal for the unperturbed background geometry $g_0$. 
	\item expand the area functional using (\ref{eq:metricpert}) and (\ref{eq:positionpert}) 
	\item extremize the area functional with respect to $x_1,x_2,...$
\end{enumerate} 
The extremization procedure yields ``equations of motion'' for $x_1,x_2,...$, which can be solved in certain cases. These ``equations of motion'' correspond with the condition that the expansions of an extremal surface vanish \cite{Hubeny:2007xt}. The solutions for $x_1,x_2,...$ can be substituted back into the area functional.

In sections (\ref{sec:gauge}) and (\ref{sec:higher}) we also show that the extremization problem at any order in $\lambda$ can be reduced to that of order one, by using gauge transformations. Furthermore, in section (\ref{sec:hwgauge}) we derive solutions for a series of diffeomorphisms that bring the metric into the Hollands-Wald gauge.

\subsection{Results}
The extremization procedure described in section (\ref{sec:var}) yields ``equations of motion'' for the ``shifts'' $x_1,x_2,\dots$ (\ref{eq:positionpert}). These ``equations of motion'' correspond to the requirement that the expansions $\K$ where $i=0,1$ labels the normal vectors $\N$, vanish at all orders in $\lambda$:
\begin{align}
\left.\frac{d\K}{d\lambda}\right|_{\lambda=0} &= 0 &\Leftrightarrow \textmd{e.o.m. for }x_1^a \label{eq:eom1}\\
\left.\frac{d^2\K}{d\lambda^2}\right|_{\lambda=0} &= 0 &\Leftrightarrow \textmd{e.o.m. for }x_2^a \label{eq:eom2}\\
\dots &\dots \dots &\dots\nonumber 
\end{align}
The first equation (\ref{eq:eom1}) can be simplified by choosing $x_1$ to be perpendicular to the original surface: $x_{1\parallel}^a=0$. In this article, we focus mostly on metric perturbations of $\textmd{AdS}_{d+1}$, and ball-shaped sub-regions of the CFT,\footnote{In this case, the HRRT-surfaces are totally geodesic; their extrinsic curvatures vanish completely.} for which the equation of motion of $x_1$ can be solved by using a Green's function:
\begin{align}
x_1^a &= \sum_i \left(\int_{\tilde{B}} d^{d-1}\beta \sqrt{h} \  G_{\tilde{B}}(\alpha,\beta) \ \delta^{1}_{g^1}\K(\beta)\right) \N^a_0(\alpha) 
\end{align} 
where $G_{\tilde{B}}(\alpha,\beta)$ is a Green's function on the original HRRT-surface $\tilde{B}$, which has hyperbolic geometry, satisfying\footnote{A Green's function is generally not unique without specifying the appropriate boundary conditions; in this case we require the response function to vanish at the boundary. For a discussion of Green's functions on hyperbolic space, see for example \cite{grellier:hal-00022000,stoll2016harmonic,Haba:2007ay}}
\begin{align}\label{eq:greens}
\left(\square_{\tilde{B}}-(d-1)\right)G_{\tilde{B}}(\alpha,\beta) = \frac{1}{\sqrt{h_0}}\delta^{d-1}(\vec{\alpha}-\vec{\beta}),
\vspace{-10mm}
\end{align}  
and $\delta^1_{g^1}\K$ is the first order metric perturbation of the expansions $\K$, evaluated at $\delta g_{ab} = g^1_{ab}$ (for an explicit expression, see equation \ref{eq:expmet}). 

As described in sections (\ref{sec:gauge}) and (\ref{sec:higher}), a series of diffeomorphisms can be used to reduce the extremization problem for $x_2,x_3,\dots$ to a problem of the same complexity as the extremization problem for $x_1$. For $x_2$, this diffeomorphism is generated by the vector field $V^a = -x_1^a$, which basically ``reverses'' the shift at order $\lambda$, and the solution for $x_2$ is given by 
\begin{align}\label{eq:2ordersol}
x_2^a &= -\frac{1}{2}x_1^b\partial_bx_1^a  \\&+\sum_i \left(\int_{\tilde{B}} d^{d-1}\beta \sqrt{h} \  G_{\tilde{B}}(\alpha,\beta) \ \left(\delta^{2}_{\bar{g}^1}\K(\beta)+\delta^{1}_{\bar{g}^2}\K(\beta)\right)\right) \N^a_0(\alpha)
\\ &\textmd{with }
\bar{g}^1_{ab} = g^1_{ab} - \mathcal{L}_{x_1}g^0_{ab} \hspace{1cm}\textmd{and}\hspace{0.3cm}
\bar{g}^2_{ab} = g^2_{ab} - \mathcal{L}_{x_1}g^1_{ab}+\frac{1}{2}\mathcal{L}_{x_1}\mathcal{L}_{x_1}g^0_{ab}.
\end{align}
Similar results can be obtained for $x_3,x_4,\dots$ et cetera. 

The solutions for $x_1,x_2,...$ can be substituted back into the area functional: this provides an expansion of the area functional in terms of integrals over the original HRRT-surface $\tilde{B}$ (for ball-shaped boundary sub-regions and the $\textmd{AdS}_{d+1}$ background geometry): 
\begin{align}\label{eq:areaexp}
A(\lambda) &= A_0 + \lambda A_1+ \lambda^2 A_2 + \dots, \end{align}
where the first non-trivial term, at order $\lambda^2$, is given by:
\begin{align}
A_2 &= \frac{1}{2}\int d^{d-1}\alpha \sqrt{h_0}\left(2g^2_{ab}\frac{\delta \sqrt{h_0}}{\delta g^0_{ab}}+g^1_{ab}g^1_{cd}\frac{\delta^2\sqrt{h_0}}{\delta g^0_{ab}\delta g^0_{cd}}\right) \label{eq:2ndorderexp}\\
&+\frac{1}{2}\sum_{i=0,1}\int d^{d-1}\alpha \sqrt{h_0} \int d^{d-1}\beta \sqrt{h_0} \  G_{\tilde{B}}(\alpha,\beta)  \left(\delta^1_{g^1}\K(\alpha)\right)\left(\delta^1_{g^1}\K(\beta)\right). \label{eq:2ndordershift}
\end{align}  
More generally, the term at order $n$ in the expansion of the extremal surface area (\ref{eq:areaexp}) has the structure
\begin{align}
A_n &=
 \left(\int_{\tilde{B}}(\cdots)+\dots+\overbrace{\int_{\tilde{B}}\dots\int_{\tilde{B}}}^{\textmd{n integrals}}\overbrace{(\hspace{1cm}\cdots \hspace{1cm})}^{\textmd{n-1  Green's functions}}\right) 
\end{align}
The methods detailed below can be used in an iterative procedure, in order to find a series of diffeomorphism generating vector fields that enforce the Hollands-Wald gauge. At first order, a solution for the diffeomorphism generating vector field is given by
\begin{align}
V^a_1 &= S_1^a + K_1^a,
\end{align}
where $S$ stands for ``shift'' and $K$ for ``Killing'', as $S_1^a = -x_1^a$ reverses the shift $x_1$, such that the coordinate position of the extremal surface is unchanged, and 
\begin{align}
K_1^a &= -\frac{1}{2\pi}\tilde{g}^1_{ab}\xi^b+\frac{1}{8\pi^2}\epsilon_{ab}\tilde{g}_1^{bc}\xi_c, \hspace{1cm}\tilde{g}^1_{ab} = g^1_{ab}+\mathcal{L}_{S_1}g^0_{ab},
\end{align} 
enforces that the Killing equation is still satisfied at the extremal surface $\tilde{B}$. Here, $\xi$ is the Killing vector, for which $\tilde{B}$ is the Killing horizon \cite{Faulkner:2013ica}. Note that $K_1$ vanishes at the extremal surface $\tilde{B}$, because the Killing vector $\xi$ vanishes there. Details can be found in section (\ref{sec:hwgauge}), where we outline a procedure to construct explicit solutions for the $V_i$ at higher orders of $\lambda$. 

\section{Variational Method}\label{sec:var}
In this section, we first consider the area functional as a function of an arbitrary shift of the form (\ref{eq:positionpert}) --this is generally not an extremal surface--, and expand it in orders of $\lambda$. Extremization of the area functional with respect to $x_1,x_2,\dots$ (\ref{eq:positionpert}) yields the ``equations of motion'' (\ref{eq:eom1},\ref{eq:eom2}), that can be solved in certain cases.
\\
\\
First, we construct a basis of normal vectors, that remains orthonormal for non-zero $\lambda$, given an arbitrary shift of the position of the surface (\ref{eq:positionpert}):
\begin{align}\label{eq:normpert}
\N_{a}(\lambda) &= \N_{a} + \Delta \N_a &= \N^0_a + \lambda  \N^1_a + \lambda^2 \N^2_a+\dots 
\end{align}
where $i=0,1$ labels the two normal vectors (one timelike, one spacelike), $a$ is a regular spacetime index and the right superscript $(0,1,2,...)$ represents the order in $\lambda$ (see appendix \ref{app:notation} for notation).
The area functional now depends on $\lambda$ via: 
\begin{enumerate}
	\item the perturbed metric (\ref{eq:metricpert})
	\item the change of the position (\ref{eq:positionpert})
	\item the normal vectors (\ref{eq:normpert}) associated to (\ref{eq:metricpert}) and (\ref{eq:positionpert})
\end{enumerate}
since 
\begin{align}\label{eq:area}
A(\tilde{B}) &= \int_{\tilde{B}} d^{d-1}\alpha \sqrt{h}  \ \ \ \textmd{where} \ h_{ab} = g_{ab} + \sum_i (-)^i \  \N_a \N_b \  \textmd{is the induced metric.}
\end{align}
The HRRT-surface depends on $\lambda$ via each point of the (original) surface:
\begin{align}\label{eq:localvar}
\frac{d}{d\lambda} &= \int d^{d-1}\alpha \left\{\frac{\partial x^a}{\partial\lambda} \frac{\delta}{\delta x^a} + \frac{\partial g_{ab}}{\partial\lambda} \frac{\delta}{\delta g_{ab}}+\sum_{i=0,1}\frac{\partial \N_a}{\partial \lambda}\frac{\delta}{\delta \N_a}\right\}.
\end{align}
In what follows, the geometric flow equation will provide a key simplification:
\begin{align}\label{eq:geoflow}
\frac{\delta A}{\delta x^a} &= \sum_i \sqrt{h} \ \K\N_a.
\end{align}
\subsection{Normal Vectors}\label{sec:normal}
First, we construct the normal vectors (\ref{eq:normpert}) for the arbitrarily shifted surface (\ref{eq:positionpert}), in the perturbed geometry (\ref{eq:metricpert}).\footnote{The normal vectors $\N^0$ are defined at the surface $\tilde{B}$, where they are orthogonal and have unit norm, and can --in principle-- be extended away from $\tilde{B}$ in an arbitrary way. For convenience, we require that the $\N^0$ remain orthonormal in some neighborhood of $\tilde{B}$.} Here we state the result for the $\N^1$; for a full derivation see appendix (\ref{app:normal}).
\\
\\
The \emph{tangent} component $\N^1_{\parallel a}$ (\ref{eq:normpert}) is determined by the orthogonality condition:
\begin{align}\label{eq:n1tangent}
\N^1_{\parallel a} &= -h_a^{0 \ b}\left(\N^0_c\nabla_bx_1^c+x_1^c\nabla_c \N^0_b\right)
\end{align}
The \emph{perpendicular} component $\N^1_{\perp a}$ (\ref{eq:normpert}) is determined by requiring unit norm and orthogonality between the $\N_a,i=0,1$:
\begin{align}\label{eq:n1perp}
\N^1_{\perp a} &= \frac{1}{2}\sum_j \left(g_1^{bc} \ \N^0_b \  \Nj^0_c\right)\Nj^0_a.
\end{align} 
Note that the perpendicular component $\N_{\perp}^1$ (\ref{eq:n1perp}) only depends on the metric perturbation  $g^1$ and \emph{not} on the shift $x_1$.\footnote{For this reason, we will include the terms involving $\N_{\perp}$ in the $\delta^1_g \K$, the first order metric perturbation of the expansions $\K$. This is just a matter of ``book keeping''.}

\subsection{Expansion of the Area Functional to First Order}
The first $\lambda$-derivative (\ref{eq:localvar}) of the surface area $A$ (\ref{eq:area}) can be simplified by using the geometric flow equation (\ref{eq:geoflow})%\begin{align}\label{eq:geoflow}
%\frac{\delta A}{\delta x^a} &= \sum_i \sqrt{h}\K\N_a,
%\end{align}
:\footnote{Note that $\frac{\delta \sqrt{h}}{\delta \N_a} = 0$ at \emph{all} orders in $\lambda$: $\frac{\delta \sqrt{h}}{\delta \N_a} = \pm \sqrt{h}h^{ab}\N_b = 0$, because the induced metric $h$ projects onto the tangent space.}
\begin{align}\label{eq:firstder}
\frac{dA}{d\lambda} &= \int_{\tilde{B}} d^{d-1}\alpha \ \left(\frac{\partial g_{ab}}{\partial \lambda}\frac{\delta \sqrt{h}}{\delta g_{ab}}+\frac{\partial x^a}{\partial \lambda} \sum_i \sqrt{h}\K \N_a\right).
\end{align}
Evaluating this formula at $\lambda=0$, using $\frac{\delta \sqrt{h}}{\delta g_{ab}} = \frac{1}{2}\sqrt{h}h^{ab}$, we find
\begin{align}
\left.\frac{dA}{d\lambda}\right|_{\lambda=0} &= \frac{1}{2}\int_{\tilde{B}} d^{d-1}\alpha \sqrt{h_0}h^{ab}g^1_{ab},
\end{align}
where we used that the expansions $\K|_{\lambda=0}=0$ vanish under the assumption that $x_0(\alpha)$ is an extremal surface for the background geometry $g^0$. We recovered a well-known result \cite{Hubeny:2007xt}: at first order in $\lambda$, the change of the surface area of an extremal surface does not depend on the shift of the surface, due to the fact that the expansions of an extremal surface vanish. 

\subsection{Expansion of the Area Functional to Second Order}
We proceed by taking one more $\lambda$-derivative (using equation \ref{eq:localvar}) of (\ref{eq:firstder}) and evaluate at $\lambda=0$:
\begin{align}
\left.\frac{d^2A}{d\lambda^2}\right|_{\lambda=0} &=&& \int_{\tilde{B}} d^{d-1}\alpha \left(2g^2_{ab}\frac{\delta \sqrt{h_0}}{\delta g^0_{ab}}+g^1_{ab}g^1_{cd}\frac{\delta^2\sqrt{h_0}}{\delta g^0_{ab}\delta g^0_{cd}}\right)
\label{eq:metric2exp} \\
&+&&\int_{\tilde{B}} d^{d-1}\alpha\left(x_1^c\nabla_c\left(g^1_{ab}\frac{\delta \sqrt{h_0}}{\delta g^0_{ab}}\right)+\sum_i\N_{c\parallel}^1 \frac{\delta}{\delta \N^0_c}\left(g^1_{ab}\frac{\delta \sqrt{h_0}}{\delta g^0_{ab}}\right)\right)\label{eq:eq} \\ 
&+&&\sum_i\int_{\tilde{B}} d^{d-1}\alpha\sqrt{h_0} \  x_1^c \N_c^0\left(\left.\frac{d\K}{d\lambda}\right|_{\lambda=0}\right) \label{eq:deltak}
\end{align}

Line (\ref{eq:deltak}) contains the first order correction to the extrinsic curvature (for arbitrary $x_1$). At first order, 
\begin{align}\label{eq:expansion}
\left.\frac{d\K}{d\lambda}\right|_{\lambda=0} &= \delta^{1}_{g^1}\K +\delta^{1}_{x^1}\K +    \sum_j\delta^{1}_{\Nj^1_{\parallel}} \K  \\&=\delta^{1}_{g^1}\K+x^c_1h_0^{ab}R^0_{dbac}\N_0^d-h_0^{ab}\nabla_a\left(h_0^{bc}(\nabla_cx_1^d)\N^0_d\right) -\K^0_{ab}h_0^{ac}\nabla_cx_1^b \nonumber
\end{align}
where the $\delta^{1}_{x^1}\K$ are the first order corrections to the expansions $\K$ due to the shift (\ref{eq:positionpert}), $\delta^{1}_{g^1}\K$ is the first order correction to the expansions $\K$ due to the change in the metric (\ref{eq:metricpert});
\begin{align}\label{eq:expmet}
\delta^{1}_{g^1}\K &= - h_0^{ab}\delta^{1}_{g^1}\Gamma^c_{ab}\N^0_c-g_1^{ab} \ {}^{i}\hspace{-1mm}K_{ab}^0, \\ \delta^{1}_{g^1}\Gamma^c_{ab} &= \frac{1}{2}g_0^{cd}\left(\nabla_{a}g^1_{bd}+\nabla_{b}g^1_{ad}-\nabla_{d}g^1_{ab}\right)
\end{align}
and $\delta^{1}_{\Nj_{\parallel}} \K$ is the first order correction due to the tangent component of the change of the normal vector $\Nj$ (for a derivation of (\ref{eq:expansion}), see appendix \ref{app:extr}).\footnote{For totally geodesic surfaces $\K^0_{ab}=0$).}
\\
Line (\ref{eq:eq}) can be rewritten as
\begin{align}\label{eq:partial}
& \sum_i\int_{\tilde{B}} d^{d-1}\alpha \sqrt{h_0}(x_1^c\N_c^0)\left(\delta^{1}_g\K \right)+\sum_i\int_{\tilde{B}} d^{d-1}\alpha \sqrt{h_0} D_a \left(g_1^{ab}\N_b^0 x_1^c\N_c^0\right), 
\end{align} 
where the second term of line (\ref{eq:partial}) is a boundary term. 
\subsection{Extremization of the Area Functional with respect to $x_1(\alpha)$}
Using some algebra and one partial integration (\ref{eq:partial}), we rewrite the terms at second order, equations (\ref{eq:metric2exp}), (\ref{eq:eq}) and (\ref{eq:deltak}), as
\begin{align}
\left.\frac{d^2A}{d\lambda^2}\right|_{\lambda=0} &=&& \int d^{d-1}\alpha \left(2g^2_{ab}\frac{\delta \sqrt{h_0}}{\delta g^0_{ab}}+g^1_{ab}g^1_{cd}\frac{\delta^2\sqrt{h_0}}{\delta g^0_{ab}\delta g^0_{cd}}\right)
\label{eq:metric2expplus} \\
&+&&\sum_i \int d^{d-1}\alpha \sqrt{h_0} \ (x_1^d\N^0_d) \left(\delta^{1}_g\K + \left.\frac{d\K}{d\lambda}\right|_{\lambda=0} \right),\label{eq:x1dep}
\end{align}
where only the second line contains terms involving $x_1$. 

Extremizing with respect to $x_1$ gives an ``equation of motion'' for $x_1$, which in turn determines $x_1$ in terms of the metric perturbation $g^1$. Using (\ref{eq:x1dep}) and (\ref{eq:expansion}) we find that the equation of motion for $x_1$ is equivalent to the requirement that the expansions vanish at first order in $\lambda$:
\begin{align}\label{eq:x1eom}
0 &= \left.\frac{d\K}{d\lambda}\right|_{\lambda=0} \\
&= \delta^{1}_{g^1}\K+x^c_1h_0^{ab}R^0_{dbac}\N_0^d-h_0^{ab}\nabla_a\left(h_0^{bc}(\nabla_cx_1^d)\N^0_d\right) -\K_{ab}h_0^{ac}\nabla_cx_1^b
\end{align}
This is a well-known result: the expansions of an extremal surface must vanish. This requirement determines $x_1,x_2,...$, for which the ``equations of motion'' can be found by expanding the $\K$ in orders of $\lambda$, and requiring these the terms in this expansion to vanish. The ``equation of motion'' for $x_1$ (\ref{eq:x1eom}) also appears as equation (47) of \cite{LashkariVanRaamsdonk2016} or equation (2.25) of \cite{Faulkner:2017tkh}. \emph{Assuming} that $x_1$ solves equation (\ref{eq:x1eom}), $A_2$ (\ref{eq:metric2expplus},\ref{eq:x1dep}) becomes
\begin{align}
A_2 &= \frac{1}{2}\int d^{d-1}\alpha \left(2g^2_{ab}\frac{\delta \sqrt{h_0}}{\delta g^0_{ab}}+g^1_{ab}g^1_{cd}\frac{\delta^2\sqrt{h_0}}{\delta g^0_{ab}\delta g^0_{cd}}\right)+\frac{1}{2}\sum_i \int d^{d-1}\alpha \sqrt{h_0} \ (x_1\cdot\N^0) \left(\delta^{1}_g\K \right).\nonumber
\end{align}
We can make connection to a problem that has been described in the mathematics literature: consider a minimal surface, and shift the position of that surface by $\hat{n}f$, where $\hat{n}$ is the normal vector and $f$ a function; what is the change in the area? In our set-up, this question corresponds to setting $\Delta g = g^1 = g^2 = \dots = 0$, and writing $x_1^a = \sum_i f^1_i\N^a_0$, which gives
\begin{align}
A_2 &= \frac{1}{2}\sum_i\sum_j \int_{\tilde{B}} d^{d-1}\alpha \sqrt{h_0} f^1_i \left( h_0^{ab}R^0_{cbad}\N_0^c\Nj^d_0-\delta^{ij}\square_{\tilde{B}} -\K_{ab}\Kj^{ab} \right)f^1_j.
\end{align}
The term in brackets is the Jacobi or stability operator.\footnote{Usually, this formula is stated for codimension-one surfaces. Some background material can be found in \cite{colding1999minimal} or \cite{calegari2007foliations}, for example.}

\subsection{Solution for $x_1(\alpha)$}
In this subsection, we will solve the equation of motion for $x_1$ (\ref{eq:x1eom}), for perturbations of $\textmd{AdS}_{d+1}$, and ball-shaped boundary sub-regions. The ``shift'' $x_1$ can always be decomposed in a tangential part and a perpendicular part:
\begin{align}
x^{1 \ a}_{\parallel} &= h_0^{ab}x^1_b, \hspace{1cm}
x^{1 \ a}_{\perp} = (g_0^{ab}-h_0^{ab})x^1_b = \sum_i f^1_i \N^a_0,
\end{align}
which defines the set of functions $f^1_i = (-)^i x_1\cdot \N$. 
A tangential shift corresponds to a re-parametrization of the original surface, which does not affect the surface area, and for simplicity we will consider
\vspace{-5mm}
\begin{align}\label{eq:x1normal}
x^1_{\parallel \ a} &= 0.
\end{align}
We will restrict the discussion to the $\textmd{AdS}_{d+1}$ background geometry, for which $
R^0_{abcd} = -(g^0_{ac}g^0_{bd}-g^0_{ad}g^0_{bc})
$, and HRRT-surfaces that end on ball-shaped boundary sub-regions. In $\textmd{AdS}_{d+1}$, HRRT-surfaces that end on ball-shaped boundary sub-regions are totally geodesic ($\K^0_{ab} = 0$), and have hyperbolic geometry ($\tilde{B}\sim \mathbb{H}^{d-1}$). In this case, the equations for the $f^1_i$ decouple, and equation (\ref{eq:x1eom}) can be written as an inhomogeneous wave equation:
\begin{align}\label{eq:wave}
0& = \left(\square_{\tilde{B}}-(d-1)\right)f^1_i\pm \delta^{1}_{g^1}\K  
\end{align}
where the sign is $\pm$ for the timelike / spacelike normal vector respectively. \\ \newpage
Equation (\ref{eq:wave}) can be solved with a Green's function for $\mathbb{H}^{d-1}$, satisfying equation (\ref{eq:greens}):\footnote{A Green's function is generally not unique without specifying the appropriate boundary conditions; in this case we require the response function to vanish at the boundary. For a discussion of Green's functions on hyperbolic space, see for example \cite{grellier:hal-00022000,stoll2016harmonic,Haba:2007ay}}
\begin{align}\label{eq:x1sol}
f^1_i(x(\alpha)) &= \int_{\tilde{B}} d^{d-1}\beta \sqrt{h} \  G_{\tilde{B}}(\alpha,\beta) \ \delta^{1}_{g^1}\K(\beta) \\
x^a_1(\alpha) &= \sum_i \left(\int_{\tilde{B}} d^{d-1}\beta \sqrt{h} \  G_{\tilde{B}}(\alpha,\beta) \ \delta^{1}_{g^1}\K(\beta)\right) \N^a_0(\alpha).
\end{align}
The terms at second order in the area of the HRRT-surface (\ref{eq:metric2expplus},\ref{eq:x1dep}) can be written as\footnote{Note: in the Hollands-Wald gauge, the term in line (\ref{eq:shiftdep}) vanishes, but it's non-zero in a general gauge.}
\begin{align}
\left.\frac{d^2A}{d\lambda^2}\right|_{\lambda=0} &= \int_{\tilde{B}} d^{d-1}\alpha \left(2g^2_{ab}\frac{\delta \sqrt{h_0}}{\delta g^0_{ab}}+g^1_{ab}g^1_{cd}\frac{\delta^2\sqrt{h_0}}{\delta g^0_{ab}\delta g^0_{cd}}\right)
\label{eq:shiftindep} \\
&+\sum_i\int_{\tilde{B}}d^{d-1}\alpha\sqrt{h_0}\int_{\tilde{B}} d^{d-1}\beta\sqrt{h_0}  G_{\tilde{B}}(\alpha,\beta)  \delta^{1}_{g^1}\K(\alpha)  \delta^{1}_{g^1}\K(\beta).\label{eq:shiftdep}
\end{align}

\subsection{An Example: the Planar Black Hole Geometry}\label{sec:example}
In this subsection we will briefly apply our formulas to a specific example, which was previously studied in \cite{LashkariVanRaamsdonk2016}: the perturbation from the $\textmd{AdS}_{2+1}$ geometry to the planar black hole geometry, for which the metric is given by:
\begin{align}
ds^2 &= \frac{1}{z^2}\left(dz^2  + \left(1+\frac{1}{2}\lambda z^2\right)^2dx^2 -\left(1-\frac{1}{2}\lambda z^2\right)^2 dt^2\right).
\end{align}
\vspace{-2mm}
Note that one recovers the background $\textmd{AdS}_{2+1}$ metric as $\lambda \rightarrow 0$. The non-trivial components of $g^1$ and $g^2$ (see equation \ref{eq:metricpert}) are given by \begin{align}
g^1_{xx} = 1, \hspace{1cm}g^1_{tt} = 1, \hspace{1cm}g^2_{xx} = \frac{z^2}{4} \hspace{1cm}\textmd{    and} \hspace{1cm}g^2_{tt}= -\frac{z^2}{4}.
\end{align}
In appendix (\ref{app:example}) we explicitly compute $x_1$ (see equation \ref{eq:x1sol}) and $A_2$ (see equations \ref{eq:shiftindep} and \ref{eq:shiftdep}), by using the appropriate Green's function for the HRRT-surface. The HRRT-surfaces in $\textmd{AdS}_{2+1}$ are space-like geodesics, and in terms of the proper distance $s$, the Green's function is given by the solution to the Helmholtz equation with constant $c=-1$:
\begin{align}
G(s,\tilde{s}) &= -\frac{1}{2}e^{-|s-\tilde{s}|}.
\end{align}
Direct computation of $x_1$, further detailed in appendix (\ref{app:example}), gives
\begin{align}
x^a_1(\tilde{s}) &= \int_{-\infty}^{\infty} ds \    G_{\tilde{B}}(s,\tilde{s}) \ \delta^1_{g^1}{}^{S}\hspace{-1.1mm}K(s)
\cdot S^a(\tilde{s}) = \frac{1}{6}\left(\frac{z(\tilde{s})}{R}(2R^2-z(\tilde{s})^2) \right)\cdot S^a(\tilde{s}),
\end{align}
where $S$ is the space-like normal vector of the HRRT-surface (in the $t=0$ plane). In appendix (\ref{app:example}) we show that the term $A_2$ in the expansion of the area (\ref{eq:areaexp}) is equal to $
A_2 = -\frac{2}{45}R^4$. These results are consistent with \cite{LashkariVanRaamsdonk2016}.
\newpage
\section{Knight Diffeomorphisms}\label{sec:gauge}
In this section, we briefly review the action of diffeomorphisms on the coordinates $x^a$ and on two-tensors. In particular, we consider the action of a series of diffeormorphisms, at different orders of $\lambda$, on the embedding coordinates (\ref{eq:positionpert}) and the perturbed metric (\ref{eq:metricpert}). The action of diffeomorphisms on tensor fields is described in \cite{0264-9381-14-9-014}, and we will briefly review some relevant results here.

We consider a series of diffeomorphisms generated by the vector fields
\begin{align}\label{eq:series}
\lambda V_1^a, \hspace{1cm}\lambda^2 V_2^a, \hspace{1cm}\lambda^3 V_3^a,\hspace{1cm}\dots
\end{align}
In other words, we consider the diffeomorphism which takes a point a distance $\lambda$ along the integral curve of $V_1$, followed by a displacement $\lambda^2$ along the integral curve of $V_2$, et cetera. Such a diffeomorphism is called a ``knight diffeomorphism'' \cite{Sonego:1997np}.

More generally, under a diffeomorphism generated by a vector field $V$, moving a parameter distance $\beta$ along its integral curves, we have the following transformation rule for a tensor field $T$ \cite{0264-9381-14-9-014}:
\begin{align}\label{eq:ttrans}
T &\rightarrow \tilde{T} = \sum_k \frac{\beta^k}{k!}\mathcal{L}^k_{V}T.
\end{align}
Now considering the series of diffeomorphisms generated by the vector fields $\lambda V_1,\lambda^2 V_2,\dots$ (\ref{eq:series}); for the coordinate functions $x^a$ we have:
\begin{align}
x^a &\rightarrow \tilde{x}^a = x^a + \lambda V_1^a + \lambda^2 \left(V^a_2+\frac{1}{2}V_1^b\partial_bV_1^a\right)+\dots \label{eq:posgauge}
\end{align}
and a symmetric two-tensor $W$ transforms as:
\begin{align}
W_{ab} &\rightarrow \tilde{W}_{ab} = W_{ab} + \lambda \mathcal{L}_{V_1}W_{ab} + \lambda^2\left(\mathcal{L}_{V_2}W_{ab} +\frac{1}{2}\mathcal{L}_{V_1}\mathcal{L}_{V_1}W_{ab}\right)+\dots\label{eq:wtrans}
\end{align}
Using the transformation rule for symmetric two-tensors (\ref{eq:wtrans}), and sorting by orders of $\lambda$, we find that the metric perturbations $g^0,g^1,\dots$ (\ref{eq:metricpert}) transform as:

\begin{align}
g^0_{ab} &\rightarrow \tilde{g}^{0}_{ab} &&= g^0_{ab} \label{eq:met0}\\
g^1_{ab} &\rightarrow \tilde{g}^1_{ab} &&= g^1_{ab} + \mathcal{L}_{V_1}g^0_{ab} \label{eq:met1}\\
g^2_{ab} &\rightarrow \tilde{g}^2_{ab} &&= g^2_{ab} + \mathcal{L}_{V_1}g^1_{ab} + \mathcal{L}_{V_2}g^0_{ab} + \frac{1}{2}\mathcal{L}_{V_1}\mathcal{L}_{V_1}g^0_{ab}. \label{eq:met2}
\end{align}
Similarly, the embedding function (\ref{eq:positionpert}), defined by $x_0,x_1,\dots$, transforms as:
\begin{align}
x^a_0 &\rightarrow \tilde{x}^a_0 &&= x^a_0 \label{eq:pos0}\\
x^a_1 &\rightarrow \tilde{x}^a_1 &&= x^a_1 + V_1^a \label{eq:pos1}\\
x^a_2 &\rightarrow \tilde{x}^a_2 &&= x^a_2 + V_2^a + \frac{1}{2}V^b_1\partial_bV^a_1 +[V_1,x_1]^a \label{eq:pos2}
\end{align}
We will use these transformation rules in section (\ref{sec:higher}), to simplify the extremization problem at higher orders of $\lambda$.

\section{Higher Order Corrections}\label{sec:higher}
In this section we will show how a knight diffeomorphism (see section \ref{sec:gauge}) can be used to reduce the extremization problem for $x_2$ to a problem of similar complexity as the extremization problem for $x_1$. The generalization of this procedure to higher orders is trivial, but we will briefly comment on this generalization in appendix (\ref{app:generalize}).

The ``equation of motion'' for $x_2$ --equivalent to the requirement that the expansions $\K$ vanish at order $\lambda^2$-- involves terms linear and quadratic in $x_1$. Now consider the diffeomorphism generated by $V_1 = -x_1$, where $x_1$ is given by (\ref{eq:x1sol}). In the new coordinates $\tilde{x}$, we have: $\tilde{x}_1=0$ (see equation \ref{eq:pos1});\footnote{Equivalently, $\delta^1_{\tilde{g}^1}\K=0$, where $\tilde{g}^1$ is given by equation (\ref{eq:met1})} and $\tilde{\N}^1_{\parallel}=0$ (see equation \ref{eq:n1tangent}).
In these coordinates, the embedding function (\ref{eq:positionpert}) starts --by construction-- at order $\lambda^2$:\footnote{Note: $x_0^a$ is unchanged by the diffeomorphism generated by $V_1$: $x_0=\tilde{x}_0$.}
\begin{align}
\tilde{x}^a(\alpha) &= \tilde{x}^a_0(\alpha)+ \lambda^2\tilde{x}^a_2(\alpha)\dots
\end{align}
The techniques that were previously used to compute $\N^1_{\parallel}$ (see section \ref{sec:normal} and appendix \ref{app:normal}) can now be used to compute  $\tilde{\N}^2_{\parallel}$. 
The equation of motion for $x_2$ is relatively simple in these coordinates:
\begin{align}\label{eq:expansion2}
0 = \K|_{O(\lambda^2)} &= \delta^{1}_{\tilde{g}^2}\K +\delta^{2}_{\tilde{g}^1}\K+ \delta^{1}_{\tilde{x}_2}\K +   \sum_j\delta^{1}_{\tilde{\Nj}^2_{\parallel}} \K \\
&= \delta^{1}_{\tilde{g}^2}\K +\delta^{2}_{\tilde{g}^1}\K+ \tilde{x}^c_2h_0^{ab}R^0_{dbac}\N_0^d-h_0^{ab}\nabla_a\left(h_0^{bc}(\nabla_c\tilde{x}_2^d)\N^0_d\right) \\ &-\K^0_{ab}h_0^{ac}\nabla_c\tilde{x}_2^b,  \nonumber
\end{align} 
where $\tilde{g}^1$ and $\tilde{g}^2$ are given by the transformation rules (\ref{eq:met1}) and (\ref{eq:met2}).\footnote{For completeness: 
$
\tilde{g}_{ab}^1 = g^1_{ab} + \mathcal{L}_{V_1}g^0_{ab}, \hspace{1cm}\tilde{g}_{ab}^2 = g^2_{ab} + \mathcal{L}_{V_1}g^1_{ab}+\frac{1}{2}\mathcal{L}_{V_1}\mathcal{L}_{V_1}g^0_{ab}.
$ Note: at this step, we only consider the diffeomophism generated by $V_1$.}
We require $\tilde{x}_2$ to be perpendicular to the original surface (similar to condition (\ref{eq:x1normal}) on $x_1$), $\tilde{x}_{2\parallel}=0$, so that we can write $\tilde{x}_{2}$ as $
\tilde{x}_2^a = \sum_i f^2_i \N^a_0
$.
Restricting the discussion to the $\textmd{AdS}_{d+1}$ background geometry and ball-shaped boundary sub-regions, the ``equation of motion'' for $\tilde{x}_2$ becomes
\begin{align}\label{eq:x2eom}
0& = \left(\square_{\tilde{B}}-(d-1)\right)f^2_i\pm \left(\delta^{1}_{\tilde{g}^2}\K +\delta^{2}_{\tilde{g}^1}\K\right).
\end{align}
We can solve (\ref{eq:eom2}) with a Green's function on $\mathbb{H}^{d-1}$, similar to (\ref{eq:x1sol}):
\begin{align}
f^2_i &= \int_{\tilde{B}} d^{d-1}\beta \sqrt{h} \  G_{\tilde{B}}(\alpha,\beta) \ \left(\delta^{2}_{\tilde{g}^1}\K(\beta)+\delta^{1}_{\tilde{g}^2}\K(\beta)\right)
\end{align}
Using the relation between $\tilde{x}_2$ and $x_1$ and $x_2$ (see equation \ref{eq:pos2}), we find
\begin{align}\label{eq:x2sol}
x_2^a &= -\frac{1}{2}x_1^b\partial_bx_1^a  \\ &+\sum_i \left(\int_{\tilde{B}} d^{d-1}\beta \sqrt{h} \  G_{\tilde{B}}(\alpha,\beta) \ \left(\delta^{2}_{\tilde{g}^1}\K(\beta)+\delta^{1}_{\tilde{g}^2}\K(\beta)\right)\right) \N^a_0(\alpha)\nonumber
\end{align}
This procedure can be repeated for $x_3,x_4$ et cetera. We will briefly comment on this generalization in appendix (\ref{app:generalize}).

\section{The Hollands-Wald Gauge}\label{sec:hwgauge}
The Hollands-Wald gauge \cite{Hollands:2012sf} allows one to exploit the Wald formalism \cite{Wald:1993nt,Iyer:1994ys}, which greatly simplifies the analysis of HRRT-surfaces, and has been used to derive (from the CFT) the gravitational equations of motion up to second order in the metric perturbation \cite{Faulkner:2013ica,LashkariVanRaamsdonk2016,Beach:2016ocq,Faulkner:2017tkh}.  However, for some applications, it is more convenient to work in a different gauge,\footnote{For example, the Fefferman-Graham gauge \cite{fgexp,Fefferman:2007rka} is often used when expressing bulk fields in terms of boundary operators (e.g. \cite{deHaro:2000vlm,Kabat:2012hp}).} and it can be cumbersome to find  a diffeomorphism that brings the metric into the HW-gauge. In this section, we devise a method to derive a solution for this problem.

The entire section is limited to the $\textmd{AdS}_{d+1}$ background geometry and HRRT-surfaces that end on ball-shaped boundary sub-regions. In $\textmd{AdS}_{d+1}$, the HRRT-surface $\tilde{B}$ for a ball-shaped boundary sub-region $B$ is a Killing horizon for a Killing vector $\xi_{\tilde{B}}$, which vanishes at the Killing horizon: $\xi_{\tilde{B}}|_{\tilde{B}}=0$.\footnote{Note: $\xi_{\tilde{B}}$ is non-zero away from $\tilde{B}$.}  
\\ \\
The Hollands-Wald gauge is a gauge in which
\begin{enumerate}
	\item the coordinate position of the surface does not change: $\Delta x(\alpha)=0$  
	\item the Killing equation is still satisfied \emph{at the surface} $\tilde{B}$ in the perturbed geometry:\begin{align}\label{eq:killing}
	\left.\mathcal{L}_{\xi}g_{ab}\right|_{\tilde{B}} &= 0
	\end{align}
\end{enumerate}
We will slightly modify the steps of section (\ref{sec:higher}), in order to compute a series of diffeomorphism generating vector fields $\lambda V_1,\lambda^2V_2,\dots$ that bring the metric into the Hollands-Wald gauge. The idea is that we reverse the shift of the coordinate position of the surface with a diffeomorphism, as in section (\ref{sec:higher}); but in addition, we must also enforce the Killing equation (\ref{eq:killing}) at the extremal surface $\tilde{B}$.

At leading order, the shift of the coordinate position of the surface, $x^1$, can be reversed by the diffeomorphism generated by\footnote{$S$ stands for ``shift'' here.}
\begin{align}\label{eq:vx}
S_1^a &= -x_1^a
\end{align}
where $x_1$ is given in terms of $g^1$ by equation \ref{eq:x1sol}. Note that $x_1$ can be extended away from the original HRRT-surface in an arbitrary way. Under such a diffeomorphism, $g^1_{ab}$ transforms as (\ref{eq:met1})
\begin{align}\label{eq:transform1}
g^1_{ab} \rightarrow \tilde{g}^1_{ab} &= g^1_{ab}+\mathcal{L}_{S^1}g^0_{ab} = g^1_{ab} +\nabla_a S^1_b +\nabla_b S^1_a.
\end{align}
An additional gauge transformation is necessary to enforce that at first order in $\lambda$, the Killing equation (\ref{eq:killing}) is satisfied at $\tilde{B}$. 

For any (symmetric) tensor field $W_{ab}$ we can construct a diffeomorphism generating vector field $K$, which satisfies:
\begin{align}\label{eq:killingpart}
\left.\mathcal{L}_{\xi}W_{ab}\right|_{\bar{B}} + \left.\mathcal{L}_{\xi}\mathcal{L}_{K}g^0_{ab}\right|_{\bar{B}} &=0 \ \ \ \ \textmd{and} \ \ \left.K\right|_{\bar{B}}=0.
\end{align}
This equation is solved by (see appendix (\ref{app:hwgauge}) for a derivation):
\begin{align}\label{eq:gaugevector}
K_a &= -W_{ab}\xi^b+\frac{1}{8\pi^2}\epsilon_{ab}W^{bc}\xi_c,
\end{align}
where $\epsilon_{ab}$ is the antisymmetric bi-normal of the surface $\tilde{B}$. 

Combining (\ref{eq:vx}),  (\ref{eq:transform1}) and (\ref{eq:gaugevector}), we find a diffeomorphism generating vector field that enforces the Hollands-Wald gauge at order one in $\lambda$:
\begin{align}
V^a_1 &= S_1^a + K_1^a, &&\textmd{with}\label{eq:hwvector} \\
S_1^a &= -x^a_1 &&\textmd{and} \label{eq:svf}\\
K_1^a &= -\frac{1}{2\pi}\tilde{g}^1_{ab}\xi^b+\frac{1}{8\pi^2}\epsilon_{ab}\tilde{g}_1^{bc}\xi_c, &&\tilde{g}^1_{ab} = g^1_{ab}+\mathcal{L}_{S_1}g^0_{ab},\label{eq:kvf}
\end{align} 
%
%
%\begin{align}\label{eq:hwgauge1}
%V^a_1 &= -x^a_1 -\frac{1}{2\pi}\bar{g}^1_{ab}\xi^b+\frac{1}{8\pi^2}\epsilon_{ab}\bar{g}_1^{bc}\xi_c \ \ \textmd{with} \ \ \bar{g}^1_{ab} = g^1_{ab}-\mathcal{L}_{x^1}g^0_{ab} 
%\\
%&=-x^a_1 -\frac{1}{2\pi}g^1_{ab}\xi^b+\frac{1}{8\pi^2}\epsilon_{ab}g_1^{bc}\xi_c+\frac{1}{2\pi}\left(\mathcal{L}_{x_1}g^0\right)_{ab}\xi^b+\frac{1}{8\pi^2}\epsilon_{ab}\left(\mathcal{L}_{x_1}g_0\right)^{bc}\xi_c
%\end{align}
where $x_1$ is given by (\ref{eq:x1sol}). At higher orders in $\lambda$, we can also use (\ref{eq:gaugevector}) to find the diffeomorphisms that bring the metric in the HW-gauge (see appendix \ref{app:generalize}). 
\\

Returning to the example of section (\ref{sec:example}), we can compute the Hollands-Wald vector $V_1$ evaluated on the HRRT-surface $\tilde{B}$, where $K_1$ vanishes and only $S_1=-x_1$ gives a non-zero contribution. In appendix (\ref{app:example}) we rewrite the result for $V|_{\tilde{B}}$ in ``spherical coordinates'' $z = r\cos \theta$ and $x = r\sin\theta$, which allows comparison with \cite{LashkariVanRaamsdonk2016}:
\begin{align}
V|_{\tilde{B}} &= \frac{R}{6}\left(\cos^2\theta-2 \right)\partial_r.
\end{align} 
This is consistent with the result found in \cite{LashkariVanRaamsdonk2016}.\footnote{See equation 63 of \cite{LashkariVanRaamsdonk2016}.}

\section{Discussion}
\subsection*{Relative Entropy at Higher Orders}
The quantum relative entropy is closely related to the entanglement entropy, whose bulk dual is given by the HRRT-formula. In this section, we will briefly review the concept of relative entropy, and we will compare our expression for the change in the extremal surface area (\ref{eq:2ndorderexp},\ref{eq:2ndordershift}) with previous results \cite{LashkariVanRaamsdonk2016}.
The relative entropy between two states described by density matrices $\rho$ and $\rho_0$ is defined as 
\begin{align}
S(\rho|\rho_0) &= \textmd{Tr}\{\rho \log \rho \} - \textmd{Tr}\{\rho \log \rho_{0} \} = \Delta \langle H_0 \rangle -\Delta S
\end{align}
with $\Delta S = S(\rho)-S(\rho_0)$, $H_0 = -\log \rho_0$ and $\Delta \langle H_0 \rangle= \textmd{Tr}\{\Delta\rho H_0 \}$. In \cite{LashkariVanRaamsdonk2016}, the Wald formalism (\cite{Wald:1993nt}) was used to establish an expression for the holographic dual of the second order terms of the relative entropy of a ball-shaped boundary sub-region $B$, by using the equations of motion:
\begin{align}\label{eq:canon}
S_B(\rho|\rho_0) &= \lambda^2 \left(\int_{\Sigma(\tilde{B})}\omega (g^1,\mathcal{L}_{\xi}g^1) + \int_{\tilde{B}} d\rho(g^1,V)\right)+O(\lambda^3),
\end{align} 
where $\omega$ is the symplectic current, and $\rho(g^1,V)$ is a (d-1)-form given in \cite{LashkariVanRaamsdonk2016}. The ``Hollands-Wald vector'' $V$ must satisfy \cite{LashkariVanRaamsdonk2016}
\begin{align}
0 &= \left.\left(g^1_{iA}+\nabla_AV_i+\nabla_iV_A\right)\right|_{\tilde{B}} \label{eq:Veq1}\\
0 &= \left.\left(\nabla_iV\nabla^iV_A+[\nabla_i,\nabla_A]V^i+\nabla^ig^1_{iA}-\frac{1}{2}\nabla_A g^{1i}_{\ \ i}\right)\right|_{\tilde{B}} \label{eq:Veq2}
\end{align}
where $i$ indexes the longitudinal directions and $A$ the transverse directions.\footnote{These equations (\ref{eq:Veq1},\ref{eq:Veq2}) correspond to equation (\ref{eq:killingpart}) and equation (\ref{eq:x1eom}) and were solved here by taking $V$ to be perpendicular (\ref{eq:x1normal}), by invoking a Green's function on the HRRT-surface $\tilde{B}$ (\ref{eq:x1sol}).}\footnote{One of these equations was also re-casted in a different form in \cite{LashkariVanRaamsdonk2016}.} The second order contribution to the holographic dual of relative entropy (\ref{eq:canon}) is also called the ``canonical energy'' \cite{Hollands:2012sf,LashkariVanRaamsdonk2016}.

In \cite{Faulkner:2017tkh}, it was shown --by means of a CFT-computation-- that the relative entropy at second order matches the holographic result (for Einstein gravity) at second order if the two central charges $a^*$ and $C_T$ of the CFT are equal.\footnote{In fact, the central charges $a^*$ and $\tilde{C}_T$ must be equal, as defined in \cite{Faulkner:2017tkh}. These central charges can be related to the central charges $a$ and $c$ of the CFT.}

In this article, we presented a closed form expression for ``Hollands-Wald vector'' $V$, which solves (\ref{eq:Veq1},\ref{eq:Veq2}): see equation (\ref{eq:hwvector}). One finds a closed form expression for the canonical energy upon substitution of (\ref{eq:hwvector}) into equation (\ref{eq:canon}). Note: in the example of sub-section (\ref{sec:example}) and appendix (\ref{app:example}), the canonical energy is given directly by equation (\ref{eq:canen}), as the modular Hamiltonian is of order one in $\lambda$ \cite{LashkariVanRaamsdonk2016}. Assuming the HRRT-formula, the vectors $\lambda V_1,\lambda^2 V_2,\dots$ that generate the knight diffeomorphism that brings the metric into the Hollands-Wald gauge can also be explicitly computed  (see section \ref{sec:higher} and appendix \ref{app:generalize}). It would be interesting to compute the (relative) entropy (in the CFT) at higher orders in the perturbation, and to match these results to geometric results. Matching these formulas at higher orders will most certainly provide conditions -- for the CFT and its quantum states to be ``holographic''.

\subsection*{Perturbations of non-AdS Geometries}
In appendix (\ref{app:other}) we briefly discuss the application of our techniques to perturbations of more general asymptotically-AdS geometries.  For the ``shift'' $x_1^a = \sum_i f^1_i \N_0^a$, the $f^1_i$ must solve\footnote{HRRT-surfaces are codimension-two surfaces with a two-dimensional normal space, spanned by some set of orthonormal vector fields $\N$. Given the label $i$, the label $\neq$ denotes the normal direction orthogonal to $\N$.}
\begin{align}\label{eq:gen2}
0  
&= &&-\sum_{j=0,1}f^1_jR_{ab}\N^a_0\Nj^b_0+(-)^{\neq i}f^1_i\N^c_0\N^d_0 {}^{\neq i}\hspace{-1.1mm}N^a_0 {}^{\neq i}\hspace{-1.1mm}N^b_0 R_{dbac} \\& &&- h_0^{ab}\delta^{1}_g\Gamma^c_{ab}\N^0_c-g_1^{ab} \ {}^{i}\hspace{-1mm}K_{ab} -\square_{\tilde{B}}f^1_i -\sum_jf^1_j\K_{ab}\Kj^{ab}.
\end{align}
It is not obvious that these equations ($i=0,1$) can always be solved, but (\ref{eq:gen2}) simplifies under some assumptions. Assuming that the background metric solves the vacuum Einstein equations\footnote{Or by considering only the terms at leading order in $G_N$.}, and assuming that the original extremal surface lies on a constant-time slice of a static background geometry, the equations of motion for the $f^1_i,i=0,1$ decouple. If we can solve equation (\ref{eq:gen2}), then the second order correction to the area functional (see equation \ref{eq:areaexp}) is given by
\begin{align}
A_2 &=\frac{1}{2}\int_{\tilde{B}} d^{d-1}\alpha \left(2g^2_{ab}\frac{\delta \sqrt{h_0}}{\delta g^0_{ab}}+g^1_{ab}g^1_{cd}\frac{\delta^2\sqrt{h_0}}{\delta g^0_{ab}\delta g^0_{cd}}\right)+ \frac{1}{2}\sum_{i=0,1}\int_{\tilde{B}}d^{d-1}\alpha f^1_i(\alpha) \delta^1_{g^1}\K.
\end{align}
For simple cases, such as the $\textmd{AdS}_{d+1}$ and AdS-Schwarzschild background geometries, equation (\ref{eq:gen2}) is an inhomogeneous wave equation. 
 
\subsection*{Quantum Extremal Surfaces}
At leading order in $G_N$, the bulk dual of boundary entanglement entropy is given by the HRRT-formula (\ref{eq:hrrt}); this formula receives corrections at sub-leading orders \cite{Faulkner:2013ana,Dong:2017xht}:
\begin{align}\label{eq:correction1}
S &= \frac{A}{4G_N}+S_{\textmd{bulk}},
\end{align}
where $S_{\textmd{bulk}}$ is the bulk entanglement entropy across the surface that extremizes (\ref{eq:correction1}). This surface is called the ``quantum extremal surface'' \cite{Engelhardt:2014gca}.
In addition to the expansion in $\lambda$, the parameter that governs the perturbation of the state (\ref{eq:denspert}), one can consider an expansion $G_N$ as well. In particular, for the quantum extremal surface, the shifts $x_1,x_2,\dots$ (\ref{eq:x1sol},\ref{eq:x2sol}) will receive corrections  at sub-leading orders of $G_N$.\footnote{For example, $S_{\textmd{bulk}}$ depends linearly on $x_1$ \cite{Dong:2017xht}, and the equation of motion (\ref{eq:x1eom}) receives an additional ``source term'' at sub-leading order of $G_N$. This additional term involves a two-point function of the bulk modular Hamiltonian and another operator defined in \cite{Dong:2017xht}.} If there is no asymmetry between the ``inside'' and ``outside'' regions of the \emph{extremal} surface, there are no corrections \cite{Dong:2017xht} at sub-leading order of $G_N$. Our results thus hold at subleading order in $G_N$ for (coherent) states for which only the stress tensor is sourced. 

%\subsection*{Generalized Theories of Gravity}
%The methods developed in this article can --in principle-- be extended to generalized theories of gravity. For generalized theories of gravity, the bulk dual of entanglement entropy is thought to be dual to the ``generalized area'', given by the appropriate Wald-functional and some additional terms involving the extrinsic curvature \cite{Hung:2014npa,Dong:2013qoa}. One can still use expression (\ref{eq:localvar}) for the $\lambda$-derivative, and extremize the holographic entanglement entropy functional order-by-order.\footnote{For example, the holographic entanglement entropy functional for Gauss-Bonnet gravity involves an integral of the intrinsic curvature. The intrinsic curvature of the HRRT-surface can be expressed in terms of its extrinsic curvatures, the Riemann and Ricci tensors and the scalar curvature, by using the Gauss-Codazzi equations. Expanding these quantities in terms of $g^1$, $x^1$ and $\N^1_{\parallel}$ is more complicated than expanding the area functional (\ref{eq:area}), but can in principle be done.} We will leave this for future work. 

\section*{Acknowledgements}
I would like to thank Aitor Lewkowycz and Felix Haehl for useful discussions. This work was supported by The Netherlands Organisation for Scientific Research (NWO).
\appendix
\addtocontents{toc}{\protect\setcounter{tocdepth}{0}}
\section{Notation \& Conventions}\label{app:notation}
Surfaces:
\begin{align}
B &\textmd{ boundary sub-region} \\
\tilde{B} &\textmd{ bulk HRRT-surface for $B$, parametrized by }\{\alpha\}
\end{align}
For ball-shaped boundary sub-regions and the AdS background geometry we have
\begin{align}
\tilde{B} \sim \mathbb{H}^{d-1}.
\end{align}
The normal vectors of the bulk HRRT-surface $\tilde{B}$:
\begin{align}
\N_a 
\end{align}
where
\begin{itemize}
\item[$i$] labels the normal vectors: for a codimension-$k$ surface we have $i=0,...,k-1$
\item[$a$] is the normal bulk space-time index
\end{itemize}
We consider normal vectors that remain orthonormal in some neighborhood of $\tilde{B}$; otherwise, there are no restrictions away from $\tilde{B}$.

\begin{itemize}
	\renewcommand{\labelitemi}{\scriptsize$\blacksquare$}
	\item Numeric sub or superscripts denote the order in the small parameter $\lambda$
	\item $h_{ab}$ denotes the induced metric of $\tilde{B}$:
	\begin{align}
	h_{\mu\nu} &= \frac{\partial x^a}{\partial \alpha^{\mu}}\frac{\partial x^b}{\partial \alpha^{\nu}}g_{ab}, \hspace{1cm}
	h_{ab} = g_{ab}+\sum_i (-)^i \ \N_a\N_b
	\end{align}
	\item The expansions $\K$ are given by:
	\begin{align}
	\K_{ab} = h_a^{c}\nabla_c\N_b, \K = h^{ab}\K_{ab}
	\end{align}
	\item indices are always raised and lowered with the background metric $g^0_{ab}$
	\item the covariant derivative $\nabla_a$ always refers to the covariant derivative for the background metric $g^0_{ab}$
\end{itemize}
\textbf{Variations:} $\delta^i_{g^j}$ refers to the $i$-th order variation of a quantity with respect to the metric, evaluated at $\delta g_{ab} = g^j_{ab}$. For example,
\begin{align}
\delta^1_{g_1}\K &= - h_0^{ab}\delta^{1}_{g^1}\Gamma^c_{ab}\N^0_c-g_1^{ab} \ {}^{i}\hspace{-1mm}K_{ab},\hspace{5mm}
\delta^{1}_g\Gamma^c_{ab} = \frac{1}{2}g_0^{cd}\left(\nabla_{a}g^1_{bd}+\nabla_{b}g^1_{ad}-\nabla_{d}g^1_{ab}\right).\nonumber
\end{align}

\section{Change of the normal vector}\label{app:normal}
The vectors tangent to the surface given by the embedding equation (\ref{eq:positionpert}) are given by
\begin{align}
\frac{dx^a}{d\alpha^{\mu}} &= \frac{dx_0^a}{d\alpha^{\mu}}+\lambda \frac{dx_1^a}{d\alpha^{\mu}}+\dots
\end{align}
At the new position (\ref{eq:positionpert}), the $\N$ (\ref{eq:normpert}) can be expanded as:
\begin{align}
\N_a(x(\alpha)) &= \N_a^0(x(\alpha))+\lambda \N_a^1 (x(\alpha))+\dots \\
&= \N_a^0(x(\alpha))+\lambda \left(\N_a^1 (x_0(\alpha))+x_1^c\partial_c\N^0_a(x_0(\alpha))\right)+\dots
\end{align}
\textbf{Orthogonality} determines the tangent component of $n_1$:
\begin{align}
0&= \frac{dx^a}{d\xi^{\alpha}}\N_a(x(\alpha)) \\
&= \left(\frac{dx_0^a}{d\alpha^{\mu}}+\lambda \frac{dx_1^a}{d\alpha^{\mu}}+\dots\right)\left(\N_a^0(x_0(\alpha))+\lambda \left(\N_a^1 (x_0(\alpha))+x_1^c\partial_c\N^0_a(x_0(\alpha))\right)+\dots\right) \nonumber\\
&= \frac{dx_0^a}{d\alpha^{\mu}}\N_a^0(x_0(\alpha))+\lambda \left\{\frac{dx_0^a}{d\alpha^{\mu}}\left(\N_a^1 (x_0(\alpha))+x_1^c\partial_c\N^0_a(x_0(\alpha))\right)+\frac{dx_1^a}{d\alpha^{\mu}}\N_a^0(x_0(\alpha))\right\}+\dots\nonumber
\end{align}
The term at order $O(\lambda^0)$ vanishes by requiring the $\N_0$ to be normal to the original surface. At order $O(\lambda)$, the term in brackets must vanish. We replace the tangent vector $\frac{dx_0^a}{d\alpha^{\mu}}$ by the induced metric $h_0^{ab}$, since the term in brackets must vanish for all the tangent vectors. This yields
\begin{align}
0&= h_0^{ab}\left(\N_a^1 +x_1^c\partial_c\N^0_a\right)+h_0^{bc}(\partial_cx_1^a)\N_a^0 \\
&= h_0^{ab}\left(\N_a^1 +x_1^c\nabla_c\N^0_a\right)+h_0^{bc}(\nabla_cx_1^a)\N_a^0
\end{align}
This equation determines the tangent component of the $\N_1$:
\begin{align}
\N_{1 \parallel}^a &= -h_0^{ab}\left( x_1^c\nabla_c\N^0_a+(\nabla_bx_1^c)\N_c^0\right)
\end{align}
For $x_1^a = \sum_if^1_i\N_0^a$ 
\begin{align}
\N_{1 \parallel}^a &= -\sum_jh_0^{ab}f^1_j \Nj_0^c\nabla_c\N^0_b- h_0^{ab}\nabla_bf^1_i
\end{align}
\textbf{Normality} of $\N_a$ completely determines the normal component of $\N^1_a$:
\begin{align}
\N_a\Nj_bg^{ab} &=  \left(\N_a^0+\lambda \N_a^1+\dots\right)\left(\Nj_b^0+\lambda \Nj_b^1+\dots\right)\left(g_0^{ab}-\lambda g_1^{ab}+\dots\right) \\ &+ \lambda x_1^c\partial_c \left(\N^0_a\Nj^0_bg_0^{ab}\right) \\
&= 1 + \lambda \left\{\N^0_a\Nj^1_bg_0^{ab}+\N^1_a\Nj^0_bg_0^{ab}-\N^0_a\Nj^0_bg_1^{ab}\right\}+\dots
\end{align} 
The quantity between brackets must vanish for $i=j$ and for $i\neq j$; note that the tangent component of $\N^1$ does not affect the normalization at order one in $\lambda$. \emph{A} solution is given by:\footnote{This is \emph{a} solution, as one can add an arbitrary rotation in the normal space.}
\begin{align}
\N_a^{1\perp} &= \frac{1}{2}\sum_{j}\left(\N^0_b\Nj^0_cg_1^{bc}\right)\Nj^0_a
\end{align}
Now $\N^1$ is fully determined in terms of $x_1$ and $g^1$:
\begin{align}
\N^1_a &= \N^{1\perp}_a+\N^{1\parallel}_{a}\\
 &= \frac{1}{2}\sum_{j}\left(\N^0_b\Nj^0_cg_1^{bc}\right)\Nj^0_a-h_0^{ab}\left( x_1^c\nabla_c\N^0_a+(\nabla_bx_1^c)\N_c^0\right)
\end{align}

\section{Change of the Extrinsic Curvature}\label{app:extr}
The expansions of the surface $x_0(\alpha)$ in the background metric $g^0$ vanish by assumption. In the perturbed geometry $g_{ab}(\lambda)$ (\ref{eq:metricpert}) the surface with embedding function $x^a_0(\alpha)$ (\ref{eq:positionpert}) has a non-vanishing extrinsic curvatures, which can be expanded in orders of $\lambda$. Taking into account the change in the metric, the change in position and the change of the normal vectors, we have at first order:
\begin{align}
\left.\frac{d\K}{d\lambda}\right|_{\lambda=0} &= \delta^{1}_{x^1}\K + \delta^{1}_{g^1}\K +  \sum_j\delta^{1}_{\Nj^1_{\parallel}} \K 
\end{align}
with 
\begin{align}
\delta^{1}_{x^1}\K &= x_1^c\nabla_c \K^0\\ &= x_1^c h_0^{ab}\nabla_c\nabla_a\N^0_b\pm x_1^c(\nabla_c\N^0_b)\N^a_0\nabla_a\N_0^b \\
\delta^{1}_{g^1}\K &= - h_0^{ab}\delta^{1}_{g^1}\Gamma^c_{ab}\N^0_c-g_1^{ab}\K^0_{ab}\\ &= \N_0^c h_0^{ab}\left(\nabla_a g^1_{bc} +\nabla_b g^1_{ac} - \nabla_c g^1_{ab}\right) -g_1^{ab}\K^0_{ab} \nonumber
\end{align}
and
\begin{align}
\sum_j\delta^{1}_{\Nj_{\parallel}} \K &= \mp x_1^c(\nabla_c\N^0_b)\N^a_0\nabla_a\N_0^b -\K^0_{ab}h_0^{ac}\nabla_cx_1^b\\ &-h_0^{ab}x^c_1\nabla_a\nabla_d\N^0_c-h_0^{ab}\nabla_a\left(h^0_{bc}(\nabla^cx^d_1)\N^0_d\right)
\end{align}
Combining all these terms, we find
\begin{align}\label{eq:kexpx}
\left.\frac{d\K}{d\lambda}\right|_{\lambda=0} &= \delta^{1}_{g^1}\K+x^c_1h_0^{ab}R^0_{dbac}\N_0^d\\ &-h_0^{ab}\nabla_a\left(h_0^{bc}(\nabla_cx_1^d)\N^0_d\right) -\K^0_{ab}h_0^{ac}\nabla_cx_1^b
\end{align}
If we take $x^a_1=\sum_{j}f^1_j\Nj_0^a$, we find:
\begin{align}\label{eq:kexpf}
\left.\frac{d\K}{d\lambda}\right|_{\lambda=0} &= \delta^{1}_{g^1}\K+\sum_j f^1_j\Nj^c_0h_0^{ab}R^0_{dbac}\N_0^d-\square_{\tilde{B}}f^1_i -\sum_jf^1_j\K^0_{ab}\Kj_0^{ab}
\end{align}
\newpage
\section{An Iterative Procedure for the Computation of $V_n$}\label{app:generalize}
In this appendix we comment on the generalization of section (\ref{sec:higher}) to arbitrary order. 

Suppose the vectors $V_1,...,V_{n-1}$ are already constructed. The action of the knight diffeomorphism generated by these vector fields on the embedding function (\ref{eq:positionpert}) and the metric (\ref{eq:metricpert}) involves a number of combinatorial factors, as described below. We will now describe how to construct $V_n$.
\begin{enumerate}
	\renewcommand{\labelitemi}{\scriptsize$\blacksquare$}
	\item define $\bar{g}^m_{ab}$, with $0 \le m \le n$ as the sum of all combinations
	\begin{align}\label{eq:bardef}
	\mathcal{L}_{V_{q_1}}\dots \mathcal{L}_{V_{q_r}} g^l_{ab},
	\end{align}
	where $q_1 \geq q_2 \geq \dots \geq q_r$, $q_1+\dots+q_r+l=m$, weighted by appropriate combinatorial pre-factors (following equation \ref{eq:ttrans}): if $N_j$ is the number of times that $V_j$ appears in (\ref{eq:bardef}), then one should multiply (\ref{eq:bardef}) by $(N_1! \dots N_{n-1}!)^{-1}$.
	\item define $\Delta_n \K$ as the sum of all combinations of the variations of the expansions $\K$,
	\begin{align}\label{eq:deltakdef}
	\delta^{1}_{\bar{g}^{q_1}}\dots \delta^{1}_{\bar{g}^{q_r}}\K,
	\end{align}
	where $\sum_{i=1}^{r} q_i = n$, weighted by appropriate combinatorial factors: if $\tilde{N}_j$ is the number of times that $\bar{g}^j$ appears in (\ref{eq:deltakdef}), then one should multiply by $\frac{n!}{\tilde{N}_1!\dots \tilde{N}_n!}$
	\item define $\Delta X_n^a$, with $0 \le m \le n$, as the sum of all combinations
	\begin{align}\label{eq:xdef}
	\mathcal{L}_{V_{q_1}}\dots \mathcal{L}_{V_{q_r}} x^a_l,
	\end{align}
	where $q_1 \geq q_2 \geq \dots \geq q_r$, $q_1+\dots+q_r+l=m$, weighted by appropriate combinatorial pre-factors: if $N_j$ is the number of times that $V_j$ appears in (\ref{eq:bardef}), then one should multiply (\ref{eq:bardef}) by $(N_1! \dots N_{n-1}!)^{-1}$.
	\item the diffeomorphism generating vector field $V_n$ is now given by
	\begin{align}
	V_n^a &= -\Delta X_n^a-x^a_n(\alpha)\\ &= -\Delta X_n^a-\sum_i \left(\int_{\tilde{B}}d^{d-1}\beta \ \sqrt{h} \  G_{\tilde{B}}(\alpha,\beta) \Delta_n \K(\beta)\right)\N^a_0(\alpha)\nonumber
	\end{align}
\end{enumerate}
For the Hollands-Wald gauge, some steps need to be modified. Lets assume that we have the vectors $V_1,\dots , V_{n-1}$ that bring the metric into the HW-gauge up to order $\lambda^{n-1}$. Steps $1-3$ are unchanged, but step $4$ is replaced by
\begin{enumerate}
	\setcounter{enumi}{3}
	\item the vector $V_n$ is given by $V_n = S_n^a + K_n^a$, with:
	\begin{align}
	S_n^a &= -\Delta X_n^a-x^a_n(\alpha)\\ &= -\Delta X_n^a-\sum_i \left(\int_{\tilde{B}}d^{d-1}\beta \ \sqrt{h} \  G_{\tilde{B}}(\alpha,\beta) \Delta_n \K(\beta)\right)\N^a_0(\alpha) 
	\end{align}
	and\begin{align}
	K_n^a &= -\frac{1}{2\pi}\tilde{g}^n_{ab}\xi^b+\frac{1}{8\pi^2}\epsilon_{ab}\tilde{g}_n^{bc}\xi_c, \hspace{1cm}\tilde{g}^n_{ab} = \bar{g}^n_{ab}+\mathcal{L}_{S_n}g^0_{ab}. \nonumber
	\end{align}
\end{enumerate}

\section{Solving for the Hollands-Wald Gauge}\label{app:hwgauge}
The problem is to find a vector field, that generates a diffeomorphism, such that one obtains a gauge in which the perturbed metric (\ref{eq:metricpert}) satisfies the Killing equation at $\tilde{B}$:
\begin{align}
\mathcal{L}_{\xi}g = 0.
\end{align}
In other words, for any (symmetric) tensor field $W_{ab}$ we can need to construct a  diffeomorphism generating vector field $K$ such that
\begin{align}\label{eq:gaugeprob}
\left.\mathcal{L}_{\xi}W_{ab}\right|_{\tilde{B}} + \left.\mathcal{L}_{\xi}\mathcal{L}_{K}g^0_{ab}\right|_{\tilde{B}} &= \left.\mathcal{L}_{\xi}W_{ab}\right|_{\tilde{B}}+ \left.\mathcal{L}_{\xi}\left(\nabla_aK_b+\nabla_bK_a\right)\right|_{\tilde{B}}=0
\end{align}
This equation is solved by
\begin{align}\label{eq:gaugesol}
K_a &= -\frac{1}{2\pi}W_{ab}\xi^b+\frac{1}{8\pi^2}\epsilon_{ab}W^{bc}\xi_c
\end{align}
This can be checked by using:
\begin{align}
\nabla_a\xi_b &= 2\pi \epsilon_{ab} &\textmd{on $\tilde{B}$, where }\epsilon_{ab}\textmd{ is the antisymmetric binormal}\\
\nabla_a \nabla_b \xi_c &= R^0_{cbad}\xi^d &\textmd{using that }\xi \textmd{ is Killing w.r.t. }g^0\label{eq:killinglemma}\\
\nabla_a \epsilon_{bc} &= 0 &\textmd{using (\ref{eq:killinglemma}) and using that $\xi|_{\tilde{B}}=0$}\label{eq:eps}
\end{align}
Note that the vector (\ref{eq:gaugesol}) vanishes at $\tilde{B}$. This solution can be used to establish the gauge in which $\mathcal{L}_{\xi}g(\lambda)=0$ at all orders in $\lambda$.\\
\\
\textbf{Proof of Solution (\ref{eq:gaugesol}) to Equation (\ref{eq:gaugeprob}):}
\begin{align}
&\left.\mathcal{L}_{\xi}W_{ab}\right|_{\tilde{B}} + \left.\mathcal{L}_{\xi}\mathcal{L}_{K}g^0_{ab}\right|_{\tilde{B}} = \left.\mathcal{L}_{\xi}W_{ab}\right|_{\tilde{B}}+ \left.\mathcal{L}_{\xi}\left(\nabla_aK_b+\nabla_bK_a\right)\right|_{\tilde{B}} \\
&=W_{ac}\nabla_b\xi^c+W_{bc}\nabla_a\xi^c +(\nabla_aK_c)\nabla_{b}\xi^c+(\nabla_cK_b)\nabla_{a}\xi^c\\&+(\nabla_bK_c)\nabla_{a}\xi^c+(\nabla_cK_a)\nabla_{b}\xi^c \nonumber\\
&= 2\pi W_{ac}\epsilon_b^c+2\pi W_{bc}\epsilon_a^c \nonumber\\
&+2\pi \left((\nabla_aK_c)\epsilon_b^c+(\nabla_cK_b)\epsilon_a^c+(\nabla_bK_c)\epsilon_a^c+(\nabla_cK_a)\epsilon_b^c\right)\nonumber
\end{align}
where still everything is evaluated at $\tilde{B}$. Now plug in solution (\ref{eq:gaugesol}), using (\ref{eq:eps}) and using that on $\tilde{B}$:
\begin{align}
\nabla_a (W_{bc}\xi^c) &= 2\pi W_{bc}\epsilon_a^c \\
\nabla_a (\epsilon_{bc}W^{cd}\xi_d) &= 2\pi  \epsilon_{bc}W^{cd}\epsilon_{ad}
\end{align}
we find:
\begin{align}
&\left.\mathcal{L}_{\xi}W_{ab}\right|_{\tilde{B}} + \left.\mathcal{L}_{\xi}\mathcal{L}_{K}g^0_{ab}\right|_{\tilde{B}} 
= 2\pi W_{ac}\epsilon_b^c+2\pi W_{bc}\epsilon_a^c \\
&-2\pi \left(W_{cd}\epsilon_a^d\epsilon_b^c+W_{bd}\epsilon_c^d\epsilon_a^c+W_{cd}\epsilon_b^d\epsilon_a^c+W_{ad}\epsilon_c^d\epsilon_b^c\right)\nonumber \\
&+\pi \left(\epsilon_{cd}W^{de}\epsilon_{ae}\epsilon_b^c+\epsilon_{bd}W^{de}\epsilon_{ce}\epsilon_a^c+\epsilon_{cd}W^{de}\epsilon_{be}\epsilon_a^c+\epsilon_{ad}W^{de}\epsilon_{ce}\epsilon_b^c\right) \nonumber\\
&= 2\pi W_{ac}\epsilon_b^c+2\pi W_{bc}\epsilon_a^c -2\pi \left(W_{cd}\epsilon_a^d\epsilon_b^c+W_{bd}\epsilon_a^d+W_{cd}\epsilon_b^d\epsilon_a^c+W_{ad}\epsilon_b^d\right)\nonumber \\
&+\pi \left(\epsilon_{bd}W^{de}\epsilon_{ae}+\epsilon_{bd}W^{de}\epsilon_{ae}+W^{de}\epsilon_{be}\epsilon_{ad}+\epsilon_{ad}W^{de}\epsilon_{be}\right)\nonumber\\
&=0 \textmd{ on $\tilde{B}$}\nonumber
\end{align}

\section{Non-AdS backgrounds}\label{app:other}
In section (\ref{sec:var}) we established that the extremization of the area functional yields the same equation of motion for $x^a_1$ as the condition that the expansions vanish at first order in $\lambda$: 
\begin{align}
\left.\frac{d\K}{d\lambda}\right|_{\lambda=0} &= \delta^{1}_{g^1}\K+x^c_1h_0^{ab}R^0_{dbac}\N_0^d-h_0^{ab}\nabla_a\left(h_0^{bc}(\nabla_cx_1^d)\N^0_d\right) -\K^0_{ab}h_0^{ac}\nabla_cx_1^b \nonumber
\end{align}
Let's proceed by choosing $x_1$ to be transverse, $x_{1\parallel}=0$, and by expanding $x_1$ on a basis of normal vectors $\{\N^0_a,i=0,1\}$:
$
x_1^a = \sum_{i=0,1} f^1_i \N_0^a
$.
The equation of motion for $x^a_1$, which holds for any boundary region and any background solution $g^0$, can now be simplified by using the symmetries of the Riemann tensor:\footnote{Using that we have a codimension-two surface}
\begin{align}
0  
&= -\sum_{j=0,1}f^1_jR_{ab}\N^a_0\Nj^b_0+(-)^{\neq i}f^1_i\N^c_0\N^d_0 {}^{\neq i}\hspace{-1.1mm}N^a_0 {}^{\neq i}\hspace{-1.1mm}N^b_0 R_{dbac} +\delta^{1}_{g^1}\K \\&-\square_{\tilde{B}}f^1_i -\sum_jf^1_j\K^0_{ab}\Kj_0^{ab}
\end{align}
If we are willing to \emph{assume} that the background metric $g^0$ is a solution of the vacuum Einstein equations, and if the original extremal surface lies on some constant-time slice of a static geometry, then the equations for the $f^1_i, i=0,1$ decouple, and we have\footnote{Here we also use that the $\N_0$ are perpendicular everywhere on $\tilde{B}$} 
\begin{align}\label{eq:gen}
0  
&= -f^1_0\frac{R-\Lambda}{2}-f^1_0\Nn^c_0\Nn^d_0 {}^{1}\hspace{-1.1mm}N^a_0 {}^{1}\hspace{-1.1mm}N^b_0 R_{dbac} +\delta^{1}_{g^1}\Kn -\square_{\tilde{B}}f^1_0, \\
0  
&= +f^1_1\frac{R-\Lambda}{2}+f^1_1\Ne^c_0\Ne^d_0 {}^{0}\hspace{-1.1mm}N^a_0 {}^{0}\hspace{-1.1mm}N^b_0 R_{dbac} +\delta^{1}_{g^1}\Ke +\square_{\tilde{B}}f^1_1-f^1_1\Ke^0_{ab}\Ke_0^{ab}.
\end{align}
Equation (\ref{eq:gen}) holds for any boundary sub-region on the constant-time slice. \\

\section{An Example: from $\textmd{AdS}_{2+1}$ to the Planar Black Hole Geometry}\label{app:example}
In this appendix we will put our formalism to work in an explicit example: the perturbation from $\textmd{AdS}_{2+1}$ to the planar black hole geometry. This example was also considered in \cite{LashkariVanRaamsdonk2016}, and our method reproduces their results.
\\

In Poincar\'e coordinates, the $\textmd{AdS}_{2+1}$ metric $g^0$ is given by:
\begin{align}\label{eq:adspoinc}
ds^2 &= \frac{1}{z^2}\left(dz^2  + dx^2 -dt^2\right).
\end{align}
The non-zero Christoffel symbols are given by:
\begin{align}\label{eq:christ}
\Gamma^{\mu}_{\nu z} = -\frac{1}{z}\delta^{\mu}_{\nu}, \hspace{1cm}\Gamma^z_{\mu\nu} = \frac{1}{z}\eta_{\mu\nu},\hspace{1cm}\Gamma^z_{zz} = -\frac{1}{z}.
\end{align}
In ``spherical coordinates'' $x=r \sin \theta$ and $z=r \cos \theta$ this metric can be written as
\begin{align}\label{eq:adsspher}
ds^2 &= \frac{1}{r^2\cos^2\theta}\left(dr^2+r^2d\theta^2-dt^2\right).
\end{align}
The ``spherical coordinates'' are used in \cite{LashkariVanRaamsdonk2016}, as they simplify certain steps in their computations.  

For a ball-shaped boundary sub-region, one can choose the coordinates such that the sub-region corresponds with the boundary interval $-R\leq x\leq R$ and $t=0$. Next, following \cite{LashkariVanRaamsdonk2016}, we consider a perturbation towards the planar black hole metric:
\begin{align}\label{eq:bhmet}
	ds^2 &= \frac{1}{z^2}\left(dz^2  + \left(1+\frac{1}{2}\lambda z^2\right)^2dx^2 -\left(1-\frac{1}{2}\lambda z^2\right)^2 dt^2\right).
\end{align}
From this metric we can determine the non-zero components of $g^1$ and $g^2$ (\ref{eq:metricpert}):
\begin{align}
g^1_{xx} &= 1, \hspace{1cm}g^1_{tt} &= 1\label{eq:g1}\\
g^2_{xx} &= \frac{z^2}{4}, \hspace{1cm}g^2_{tt}&= -\frac{z^2}{4}\label{eq:g2}
\end{align}
The HRRT-surface $\tilde{B}$ is given by $R^2 = z^2+x^2$ and $t=0$, and a set of normal vectors is given by:
\begin{align}\label{eq:normals}
T = \left[  \begin{array}{r@{\quad}cr} 
T^t  \\  
T^x \\
T^z  
\end{array}\right] = z
\left[  \begin{array}{r@{\quad}cr} 
1  \\  
0 \\
0   
\end{array}\right], \hspace{1cm}S = \left[  \begin{array}{r@{\quad}cr} 
S^t  \\  
S^x \\
S^z  
\end{array}\right] = \frac{z}{\sqrt{z^2+x^2}}
\left[  \begin{array}{r@{\quad}cr} 
0  \\  
x \\
z   
\end{array}\right].
\end{align}
We proceed with the computation of $x_1$ (see equation \ref{eq:x1sol}) and $A_2$ (see equations \ref{eq:shiftindep} and \ref{eq:shiftdep}). First we compute the first order metric correction $\delta^1_{g^1}\K$ to the expansions $\K, i=T,S$ (see equation \ref{eq:expmet}). One can check that for $g^1$ with constant components, 
\begin{align}\label{eq:christsim}
\delta^1_{g^1} \Gamma^c_{ab} &= -g_1^{cd}g^0_{de}\Gamma^d_{ab}.
\end{align}
Using equations (\ref{eq:adspoinc}), (\ref{eq:christ}), (\ref{eq:g1}), (\ref{eq:g2}), (\ref{eq:normals}), (\ref{eq:christsim}) and (\ref{eq:expmet}), we find
\begin{align}\label{eq:k}
\delta^1_{g^1}{}^{T}\hspace{-1.1mm}K = 0, \hspace{1cm} \delta^1_{g^1}{}^{S}\hspace{-1.1mm}K = -2\frac{z^3}{R^3}x^2.
\end{align}
Let $s$ be the affine parameter (proper distance) along $\tilde{B}$ defined by $x=R\tanh s$ and $z = R \textmd{ sech } s$, such that $-\infty < s < \infty$. The Green's function on the geodesic $\tilde{B}$, satisfying\footnote{Equation \ref{eq:green1} is a special case of the Helmholtz equation, of which the solution is well-known.}
\begin{align}\label{eq:helm1}
\left(\frac{d^2}{d^2s}-1\right)G(s,\tilde{s}) = \delta(s-\tilde{s}), \hspace{1cm}\lim_{s,\tilde{s}\rightarrow\pm\infty}G(s,\tilde{s})=0,
\end{align}
is given by
\begin{align}\label{eq:green1}
G(s,\tilde{s}) &= -\frac{1}{2}e^{-|s-\tilde{s}|}.
\end{align}
Now we can compute $x_1$ (see equation \ref{eq:x1sol})
\begin{align}
x^a_1(\tilde{s}) &= \int_{-\infty}^{\infty} ds \    G_{\tilde{B}}(s,\tilde{s}) \ \delta^1_{g^1}{}^{S}\hspace{-1.1mm}K(s)
\cdot S^a(\tilde{s})\end{align}
Evaluation of this integral, using equations (\ref{eq:k}) and (\ref{eq:green1}) yields:
\begin{align}\label{eq:x1s}
	x_1^a(\tilde{s}) = \left(\frac{R^2}{6} \cosh(2\tilde{s}) \textmd{sech}^3\tilde{s}\right)\cdot S^a(\tilde{s}),
\end{align}
which can also be expressed in Poincar\'e coordinates as
\begin{align}\label{eq:x1poinc}
x_1^a(z) = \frac{1}{6}\left(\frac{z}{R}(2R^2-z^2) \right)\cdot S^a(z).
\end{align}

Similarly, we can compute the ``shift dependent'' term in $A_2$ (see equation \ref{eq:shiftdep}):
\begin{align}\label{eq:shiftd}
\frac{1}{2}\int ds \int d\tilde{s}  \ G(s,\tilde{s}) \  \delta^1_{g^1}{}^{S}\hspace{-1.1mm}K(s) \  \delta^1_{g^1}{}^{S}\hspace{-1.1mm}K(\tilde{s}) = -\frac{4}{63}R^4
\end{align}
We proceed by computing the ``shift-independent'' contribution to $A_2$ (see equation \ref{eq:shiftindep}), so that we can compare our result with the ``brute force'' computation of \cite{Lashkari:2014kda,LashkariVanRaamsdonk2016}. The ``shift-independent'' term is simply the pull-back of the perturbed metric onto the original HRRT-surface:
\begin{align}
\int ds \sqrt{\frac{dx_0^a}{ds}\frac{dx^b_0}{ds}\left(g^0_{ab}+\lambda g^1_{ab}+\lambda^2g^2_{ab}+\dots\right)}.
\end{align} 
Expanding the square-root, and keeping only terms at order $\lambda^2$, we get
\begin{align}\label{eq:inte}
\lambda^2 \int ds \left(\frac{1}{2}g^2_{ab}\frac{dx_0^a}{ds}\frac{dx^b_0}{ds} -\frac{1}{8} \left(g^1_{ab}\frac{dx_0^a}{ds}\frac{dx^b_0}{ds}\right)^2\right).
\end{align} 
Using the embedding equation of the HRRT-surface, and the expressions for $g^1$ and $g^2$ (see equations \ref{eq:g1} and \ref{eq:g2}) we perform the integration: the first term in (\ref{eq:inte}) is equal to $-\frac{4}{35}R^4$ and the second term equates to $+\frac{2}{15}R^4$. Combining these result with equation (\ref{eq:shiftd}) we find
\begin{align}
A_2 = \left(-\frac{4}{63} -\frac{4}{35} +\frac{2}{15}\right)R^4 = -\frac{2}{45}R^4,
\end{align}
or equivalently
\begin{align}\label{eq:canen}
S_2 &= \frac{A_2}{4G_N} = -\frac{R^4}{90G_N},
\end{align}
which is consistent with \cite{LashkariVanRaamsdonk2016}.\footnote{See page 20, paragraph ``Comparison with relative entropy'' of \cite{LashkariVanRaamsdonk2016}.}

Finally, we discuss the vector field that generates the diffeomorphism that brings the metric in the Hollands-Wald gauge (see equation \ref{eq:hwvector}, \ref{eq:svf} and \ref{eq:kvf}), and compare our results with \cite{LashkariVanRaamsdonk2016} once more.

The vector field $V_1$ evaluated at $\tilde{B}$ is completely determined by $S_1 = -x_1$ (see equation \ref{eq:svf}), as the additional contribution $K$ (see equation \ref{eq:kvf}) only affects $V$'s derivatives; $K$ vanishes on $\tilde{B}$. Rewriting $x_1$ in ``spherical coordinates'' (see equation \ref{eq:adsspher}), we find
\begin{align}
V|_{\tilde{B}} &= S_1|_{\tilde{B}} = -x_1(\theta) \\&= -\frac{R^2}{6}\left(\cos\theta(2-\cos^2\theta) \right)\cdot S(\theta) \\
&= \frac{R}{6}\left(\cos^2\theta-2 \right)\partial_r
\end{align} 
which is consistent with \cite{LashkariVanRaamsdonk2016}.\footnote{See equation 63 of \cite{LashkariVanRaamsdonk2016}.} The derivatives of $V$ are determined by equation (\ref{eq:kvf}).
\\

%Equation (\ref{eq:gen}) is linear in $f^1_i$ and contains the Laplacian of $\tilde{B}$ acting on $f^1_i$. Assuming that we can solve equation (\ref{eq:gen}), the second order correction to the area functional is given by
%\begin{align}
%\sum_{i=0,1}\int_{\tilde{B}}d^{d-1}\alpha f^1_i(\alpha) \delta^1_{g^1}\K
%\end{align}

\bibliographystyle{ieeetr}
\bibliography{bibliography}
\end{document}